\documentclass{article}

\usepackage[nonatbib,final]{neurips_2023}
\usepackage[numbers]{natbib}

\usepackage[utf8]{inputenc} 
\usepackage[T1]{fontenc}    
\usepackage{url}            
\usepackage{booktabs}       
\usepackage{amsfonts}       
\usepackage{nicefrac}       
\usepackage{microtype}      
\usepackage{xcolor}         
\usepackage{amsmath}
\usepackage{amssymb}
\usepackage{graphicx}
\usepackage{algorithm}
\usepackage{algpseudocode}
\usepackage{bbm}
\usepackage{multirow}
\usepackage[numbers]{natbib}
\usepackage{dsfont}
\usepackage{wrapfig}
\usepackage{csquotes}
\usepackage{caption}
\usepackage{subcaption}
\usepackage{xspace}
\usepackage{wrapfig}

\usepackage{xcolor}

\usepackage[accsupp]{axessibility}  

\usepackage{array} 
\newcolumntype{H}{>{\setbox0=\hbox\bgroup}c<{\egroup}@{}}

\usepackage{subfiles} 

\usepackage{hyperref}       

\usepackage[capitalize]{cleveref}
\crefname{section}{Sec.}{Secs.}
\Crefname{section}{Section}{Sections}
\Crefname{table}{Table}{Tables}
\crefname{table}{Tab.}{Tabs.}

\DeclareMathOperator*{\argmax}{arg\,max}

\DeclareMathOperator*{\argtopk}{\mathop{\mathrm{arg\,top}}-\textit{l}\,}

\newcommand{\modelName}{ZerAuCap\xspace}
\newcommand{\mypara}[1]{\vspace{2pt}\noindent{\bf{#1}}}

\newcommand*{\affaddr}[1]{#1}

\title{Zero-shot audio captioning with audio-language model guidance and audio context keywords}

\author{%
  Leonard Salewski
  \And
  Stefan Fauth
  \And
  A.\ Sophia Koepke
  \And
  Zeynep Akata
  \AND
  \vspace{-5mm}\\
  \affaddr{
  University of Tübingen, Tübingen AI Center}\hspace{12pt}
}

\hypersetup{
pdftitle={Zero-shot audio captioning with audio-language model guidance and audio context keywords},
pdfauthor={Leonard Salewski, Stefan Fauth, A. Sophia Koepke, Zeynep Akata},
pdfkeywords={Zero-shot Audio Captioning, Large Language Models (LLM)}
}

\begin{document}

\maketitle

\begin{abstract}
    Zero-shot audio captioning aims at automatically generating descriptive textual captions for audio content without prior training for this task. Different from speech recognition which translates audio content that contains spoken language into text, audio captioning is commonly concerned with ambient sounds, or sounds produced by a human performing an action. Inspired by zero-shot image captioning methods, we propose ZerAuCap, a novel framework for summarising such general audio signals in a text caption without requiring task-specific training.
    In particular, our framework exploits a pre-trained large language model (LLM) for generating the text which is guided by a pre-trained audio-language model to produce captions that describe the audio content. Additionally, we use audio context keywords that prompt the language model to generate text that is broadly relevant to sounds.
    Our proposed framework achieves state-of-the-art results in zero-shot audio captioning on the AudioCaps and Clotho datasets. Our code is available at \url{https://github.com/ExplainableML/ZerAuCap}.
\end{abstract}

\section{Introduction}
The growing demand for seamless human-computer interaction has increased the interest in translating audio and visual information into text.
As a result, improving frameworks that automatically caption audio and visual information has gained attention. This has resulted in systems for various applications that try to improve accessibility for visually impaired or hearing impaired people~\cite{oncescu2021queryd,han2023autoad,han2023autoadiccv}.

Audio captioning aims at translating general audio signals into textual descriptions. This is different from speech recognition which transforms speech into written text. Common audio captioning methods rely on supervised learning, requiring extensive training data that spans across various audio categories. This requirement limits the adaptability of trained models to new audio contexts which is crucial as the amount and diversity of newly created audio content continue to grow.

Concurrently to our work,
\cite{shaharabany2023zero} introduced the zero-shot audio captioning task which tackles the aforementioned challenges and limitations. Their proposed approach seeks to generate textual descriptions for audio content by using guidance through an audio-text matching score that determines if the generated caption describes audible content. They use this information to iteratively optimize the key-value pairs of the LLM that generates the caption.
Different to this, our proposed \modelName framework employs a much simpler approach: We first determine a set of keywords that captures the audible events of the given soundusing a pre-trained audio-text matching network. These audio context keywords guide the text generation by being part of the input prompt for an LLM that is tasked to write a caption. Inspired by~\cite{Salewski2023ZeroshotTranslation,Su2022LanguageMC} that proposed a visually-guided zero-shot text generation framework using vision-text matching information, our framework
uses guidance from audio-text matching scores combined with the language model's word probability to select the next word out of a list of candidate words proposed by the LLM\@.
Our approach produces better results at lower computational cost than~\cite{shaharabany2023zero}.

Our \modelName framework achieves state-of-the-art results on the audio captioning benchmark AudioCaps~\cite{kim2019audiocaps}, significantly outperforming~\cite{shaharabany2023zero}. In addition, we set a new benchmark for zero-shot audio captioning on the Clotho dataset~\cite{drossos2020clotho}, outperforming our baselines.

To summarize, we make the following contributions: (1) We propose a novel framework for zero-shot audio captioning that uses a two-fold guidance approach to steer the language generation of a LLM\@. (2) Our framework achieves state-of-the-art results for zero-shot audio captioning on the AudioCaps~\cite{kim2019audiocaps} and Clotho~\cite{drossos2020clotho} datasets. (3) We show that both guiding approaches contribute to the final performance of our approach.

\section{Related Work}

\mypara{Audio captioning.} The audio captioning task consists of generating a textual description for a given sound~\cite{drossos2017automated}. This is closely related to audio-text retrieval which aims at learning to map audio and textual representations to a joint embedding space, allowing retrieval of a matching caption for a given audio snippet and vice versa~\cite{Oncescu21a,laionclap2023,slaney2002semantic,koepke2022audio,lou2022audio,Wu2022LargescaleCL}.
Audio captioning and audio-text retrieval have been popularized by the introduction of audio-text datasets, such as Clotho~\cite{drossos2020clotho}, AudioCaps~\cite{kim2019audiocaps}, Audio Caption~\cite{wu2019audio}, and WavCaps~\cite{mei2023wavcaps}.
In particular, the recurring DCASE audio captioning challenge~\cite{Dcase22} uses the Clotho dataset for benchmarking.
Many frameworks have tackled automatic audio captioning using supervision from audio-text pairs in the AudioCaps and Clotho datasets~\cite{koizumi2020audio,xu2021investigating,eren2020audio,mei2021audio,liu2021cl4ac,Kim2022ExploringTA,Gontier2021AutomatedAC,Chen2023VASTAV,Chen2023VALORVO}. Different from those works, we consider the zero-shot audio captioning setting that goes beyond zero-shot audio classification~\cite{Guzhov2021AudioclipEC} and that does not make use of task-specific training on audio captioning data. Instead, we leverage pre-trained audio-language models~\cite{Mei2023WavCapsAC} to guide the caption generation process.

\mypara{Zero-shot captioning.}
Recently, several zero-shot image captioning methods were proposed~\cite{Su2022LanguageMC,Tewel2021ZeroCapZI,Tewel2022ZeroShotVC,Zeng2022SocraticMC,Wang2022ZeroshotIC} which use CLIP~\cite{Radford2021LearningTV} to guide text generation with an LLM\@.
A first line of work optimizes hidden activations in the language model to adapt the predicted tokens towards the captioning target~\cite{Tewel2021ZeroCapZI,Tewel2022ZeroShotVC}. In contrast, our approach avoids this costly step and directly chooses the next token based on a fitness function.
A second line of work chooses the next token by introducing guidance in the decoding process~\cite{Zeng2022SocraticMC,Su2022LanguageMC}. \cite{Zeng2022SocraticMC} conditions the LLM on CLIP-detected class names and then ranks multiple sampled captions based on CLIP similarity, whilst~\cite{Su2022LanguageMC} builds a sentence token by token and selects the next token based on its CLIP similarity to the image.
In contrast,~\cite{Salewski2023ZeroshotTranslation} masks the image with the attention patterns of a VQA model and uses the same model for guiding the translation of its attention patterns into natural language.
A third line of work finetunes language models in an unsupervised way to understand CLIP text embeddings which at test time are replaced by CLIP image embeddings~\cite{Wang2022ZeroshotIC}. \cite{Li2023DeCapDC} uses the same pre-training approach and proposes a simple approach to close the modality gap in a training-free manner. Unlike the previous two approaches, our model does not rely on finetuning.
Concurrent to our work, and most closely related, \cite{shaharabany2023zero} introduced the zero-shot audio captioning task.
Different from their work we do not optimize the hidden states of the LLM, but choose the next token according to language model plausibility and audio-relevancy.

\section{Method}
In this section, we explain how our \modelName
framework (\textbf{Zer}o-shot \textbf{Au}dio \textbf{Cap}tioning) automatically generates audio captions for audio clips without task-specific training, i.e.\ in a zero-shot manner.
We use an audio-language model, pre-trained with a contrastive training objective, for two distinct purposes: \emph{zero-shot keyword selection} and \emph{audio-relevancy guiding}.
First, we select a set of top-$l$ keywords that have high similarity to the input audio clip.
These keywords are composed of very short natural language descriptions of various different audible effects, e.g.\ sounds produced by humans or other living beings, or environmental sounds.
We then provide the top keywords to the LLM to condition its subsequent generation of an audio caption on the audible concepts contained within the audio clip.
This setup exploits the world knowledge and the capability of the LLM to generate plausible sentences.
We also use the audio-language model to guide the token-by-token generation from the LLM\@. To do this, we evaluate the match of the already generated text sequence extended by the current candidate tokens with the audio clip. We then choose a token that has a high similarity to the given audio clip, whilst still being considered a likely next token by the LLM\@.

\begin{figure}[t]
    \vspace{-2ex}
    \centering
    \includegraphics[width=\linewidth]{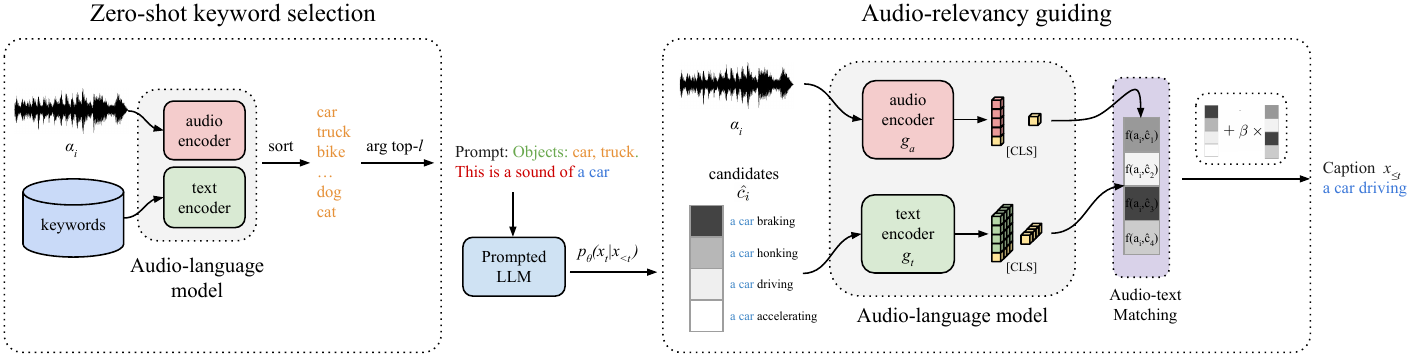}
    \caption{Our \modelName framework generates audio captions without task-specific training (zero-shot). First, we determine a set of $l$ relevant keywords that match the audio clip. An LLM is prompted with these keywords and yields the probability distribution over the next token. We evaluate the match of these candidate tokens with the input audio and choose the next token according to a weighted sum of the LLM prediction and the audio-text matching quantified by a pre-trained audio-language model.
    }%
    \label{fig:model-figure}
    \vspace{-2ex}
\end{figure}

\mypara{Zero-shot keyword selection.}
Given a list of keywords $K$, we use the pre-trained audio and text encoders of the audio-language model ($g_a$ and $g_t$) to encode the individual keywords $k$ and the audio clip $\alpha_i$. We consider the cosine similarity of the keywords and audio clip embeddings and select the top-$l$ keywords $K^*$ according to:
\begin{align}
    K^* = \argtopk_{k \in K} \text{CosSim} \left(g_a(\alpha_i),g_t(k)\right)
    = \argtopk_{k \in K} \frac{g_a(\alpha_i) \cdot g_t(k)}{\|g_a(\alpha_i)\|_2 \cdot \|g_t(k)\|_2}.
\end{align}

We compose a prompt for the pre-trained LLM which was used to generate audible candidate tokens.
Commonly, LLMs are trained to autoregressively model the probability of the next token conditioned on the previous tokens.
We exploit this capability and build a prompt which
consists of a \emph{keyword prompt} \enquote{\textsc{Objects: }}, the top-$l$ keywords $K^*$ and a \emph{default prompt} \enquote{\textsc{This is a sound of}}.
Then, the probability of the next generated token is:
\begin{equation}\label{eq:lm-prediction}
    p_{\theta}(x_t \mid
    \underbrace{x_0, \ldots, x_{b-1}}_{\text{keyword prompt}},
    \underbrace{x_{b}, \ldots, x_{d-1}}_{\text{top-$l$ keywords $K^*$}},
    \underbrace{x_{d}, \ldots, x_{h-1}}_{\text{default prompt}},
    \underbrace{x_{h}, \ldots, x_{t-1}}_{\text{autoregressive modelling}}).
\end{equation}

\mypara{Audio-relevancy guiding.}
After prompting the LLM, we re-weigh the LLM's top predicted tokens using audio-relevancy guiding. First, the LLM predicts the probability distribution over all tokens of the LLM's vocabulary. To accelerate the predictions, we then select the top-$m$ predicted tokens, which we call \emph{candidate} tokens $c_i$. As the LLM empirically assigns very small probabilities to all tokens except the top-$m$, the impact of the ignored tokens on the final weighted sum in the audio-relevancy guiding would have been negligible anyway.

We rank all candidate tokens $c_i$ by matching the currently generated sentence extended by the candidate tokens (which we will call $\hat{c}$) with the audio clip. This measures how well the generated text matches the audio clip.
The similarity $f(g_a(\alpha_i),\hat{c}_i)$ of a candidate sequence $\hat{c}_i$ with respect to the embedded audio clip $g_a(\alpha_i)$ is determined by first computing the cosine similarity $\text{CosSim}(\cdot,\cdot)$ of all possible audio-text matches,
\begin{equation}
    \begin{split}
        f(g_a(\alpha_i),\hat{c}_i)
        =
        \frac{
        e^{\kappa \cdot \text{CosSim} \left(g_a(\alpha_i), g_t(\hat{c}_i) \right)}
        }{
        \sum_{j \in 1,\ldots,k}e^{\kappa \cdot \text{CosSim} \left(g_a(\alpha_i), g_t(\hat{c}_j)\right)}
        },
    \end{split}
\end{equation}
with a scalar temperature $\kappa$.
We select the next token based on a weighted sum of the probabilities assigned to each of the candidate tokens by the LLM as well as the audio-text similarity.
For each time step $t$, the next token $x_t$ is selected according to:
\begin{equation}\label{eq:next-token}
    \begin{split}
        x_t
        =
        \argmax_{i \in 1, \ldots, m}
        \Big\{
        p_{\theta} \left(c_i \mid x_0, \ldots, x_{<t} \right) +  \beta \cdot  f_{} \left(g_a(\alpha_i), \hat{c_i} \right)
        \Big\},
    \end{split}
\end{equation}
where $\beta$ is a scalar weighting factor.
We append the selected token to the prompt and iterate this process until the language model generates a token containing a period.

\section{Experiments}

\mypara{Experimental setup.}
We adapted the 1.3B parameter version of OPT~\cite{zhang2022opt} as the LLM for all experiments unless stated otherwise. For both guiding strategies (\emph{zero-shot keyword selection} and \emph{audio-relevancy guiding}), we used the pre-trained WavCaps~\cite{Mei2023WavCapsAC} audio-language model to measure the audio-text matching.
The list of audible keywords is derived from the publicly available AudioSet class list~\cite{Gemmeke2017AudioSA}. We separated the tags for classes that contained more than one class, resulting in 614 audio keywords.
We set the number of selected keywords to $l=2$, the scalar weighting factor to $\beta=0.5$, and the number of candidates to $m=45$, using the validation set of AudioCaps~\cite{kim2019audiocaps}.
We evaluate the quality of our captions on the respective test sets of AudioCaps and Clotho~\cite{drossos2020clotho} using standard natural language generation metrics BLEU~\cite{papineni2001BLEUMethodAutomatic} (B), METEOR~\cite{banerjee2005METEORAutomaticMetric} (M), ROUGE-L~\cite{lin2004ROUGEPackageAutomatic} (RL), CIDEr~\cite{vedantam2015CIDErConsensusbased} (C), SPICE~\cite{anderson2016spice} (S), and SPIDER~\cite{liu2017spider} (Sr).

\mypara{Quantitative results.}
We present quantitative results for zero-shot audio captioning in~\Cref{tab:sota-results}. \modelName outperforms~\cite{shaharabany2023zero} by a wide margin on most metrics, setting a new state of the art on AudioCaps~\cite{kim2019audiocaps}.
Additionally, it outperforms our baseline that does not have access to any audio input on both datasets. As randomly generated texts already might have some overlap with the ground-truth captions, our baseline yields non-zero scores. Interestingly, on AudioCaps the baseline significantly outperforms the recently proposed zero-shot audio captioning model~\cite{shaharabany2023zero} in terms of the ROUGE-L score.
%
Our model always outperforms the baseline and~\cite{shaharabany2023zero} for all metrics but BLEU-4\@. This indicates that the recall of 4-grams is higher with~\cite{shaharabany2023zero}. However, this value may be skewed due to the very short captions generated by~\cite{shaharabany2023zero}.
We additionally report results from supervised methods that represent an upper bound. This comparison indicates that our model already captures significant portions of the audible content and converts them into plausible captions.

{
\renewcommand{\arraystretch}{1.2}
\begin{table*}[t]
    \centering
    \resizebox{.95\linewidth}{!}{
        \begin{tabular}{c  l H H H c c c c c c  c  H H H c c c c c c}
            \toprule
            \multicolumn{2}{c}{} & \multicolumn{9}{c}{\textbf{AudioCaps}~\cite{kim2019audiocaps}}  & & \multicolumn{9}{c}{\textbf{Clotho}~\cite{drossos2020clotho}}  \\
            \cmidrule(lr){3-11} \cmidrule(lr){13-21}
           Setting & \textbf{Framework} $\downarrow$ &  B1 & B2 & B3 & B4 & M & RL & C & S & Sr  & &  B1 & B2 & B3 & B4 & M & RL & C & S & Sr \\
           \midrule
           
            \multirow{3}*{Zero-shot} & No audio (baseline) & 18.8 & 1.1 & 0.0 & 0.0 & 4.1 & 17.8 & 0.1 & 0.0 & 0.0 &  & 18.6 & 2.8 & 0.0 & 0.0 & 3.8 & 16.6 & 0.2 & 0.1 & 0.2\\ 
            & Shaharabany et al.~\cite{shaharabany2023zero} & & & & \textbf{9.8} & 8.6 & 8.2 & 9.2 & - & - &  &  - &  - &  - &  - &  - &  - &  - & - & - \\
            \cmidrule(lr){2-21}
            & \modelName \textsubscript{OPT} (ours)  & 44.8 & 25.2 & 13.9 & 6.8 & \textbf{12.3} & \textbf{33.1} & \textbf{28.1} & \textbf{8.6} & \textbf{18.3} &  & \textbf{33.0} & \textbf{14.8} & \textbf{6.5} & \textbf{2.9} & \textbf{9.4} & \textbf{25.4} & \textbf{14.0} & \textbf{5.3} & \textbf{9.7}\\ 
            \midrule
            \multirow{2}*{Supervised} & HYU~\cite{cho2023hyu} (ensemble) & & & & - & - & - & - & - & - &  &  - &  - &  - &  20.2 &  19.7 &  42.2 &  54.1 & 14.6 & 34.3 \\
             & HTSAT-BART~\cite{Mei2023WavCapsAC}  & & &  &    28.3 &    25.0 &     50.7 &   78.7 &   18.2 & 48.5 & & & & & 16.8 & 18.4 & 38.3 & 46.2 & 13.3 & 29.7 \\
            \bottomrule
        \end{tabular}
    }
    \caption{Zero-shot audio captioning results on the AudioCaps~\cite{kim2019audiocaps} and Clotho~\cite{drossos2020clotho} datasets.}%
    \label{tab:sota-results}
\end{table*}
}

\begin{figure}
\begin{minipage}{0.45\textwidth}
    \includegraphics[width=\linewidth]{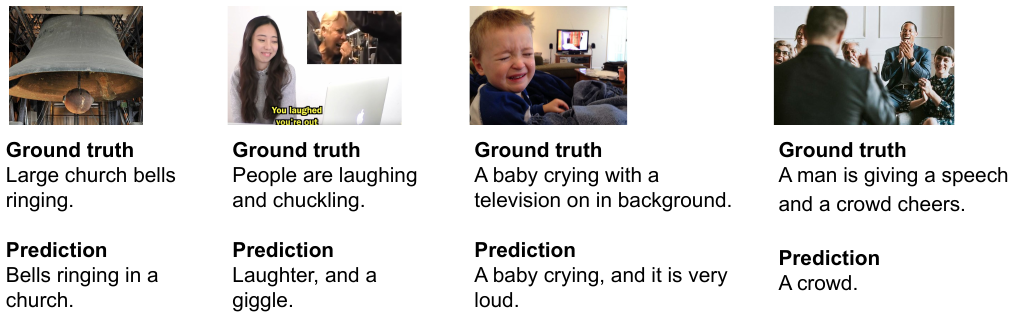}
    \captionof{figure}{Qualitative example results for \modelName. Video frames are only for illustration.}%
    \label{fig:qualitative examples}
\end{minipage}
\hfill
\begin{minipage}{0.53\textwidth}
    \resizebox{\linewidth}{!}{
        \begin{tabular}{p{2.5cm} H H H c c c c c c  H  H H H c c c c c c}
            \toprule
             & \multicolumn{9}{c}{\textbf{AudioCaps}~\cite{kim2019audiocaps}}  & & \multicolumn{9}{c}{\textbf{Clotho}~\cite{drossos2020clotho}}  \\
            \cmidrule(lr){2-10} \cmidrule(lr){12-20}
            \textbf{Ablation} $\downarrow$ &  B1 & B2 & B3 & B4 & M & RL & C & S & Sr  & &  B1 & B2 & B3 & B4 & M & RL & C & S & Sr \\
            \midrule
            No keywords & 22.3 & 7.0 & 2.7 & 0.0 & 5.9 & 20.7 & 2.5 & 2.4 & 2.4 &  & 21.9 & 6.2 & 2.0 & 0.6 & 5.4 & 18.1 & 1.4 & 1.2 & 1.3 \\ 
            No guiding with audio-relevancy & 43.9 & 24.6 & 13.5 & \multirow{2}{*}{6.3} & \multirow{2}{*}{1.2} & \multirow{2}{*}{32.8} & \multirow{2}{*}{28.1} & \multirow{2}{*}{8.2} & \multirow{2}{*}{18.1} & & \multirow{2}{*}{32.5} & \multirow{2}{*}{14.1} & \multirow{2}{*}{5.7} & \multirow{2}{*}{2.2} & \multirow{2}{*}{9.1} & \multirow{2}{*}{25.1} & \multirow{2}{*}{12.5} & \multirow{2}{*}{5.0} & \multirow{2}{*}{8.8} \\ 
            \midrule
            \modelName \textsubscript{GPT2} & 29.4 & 11.1 & 4.6 & 1.7 & 9.4 & 23.5 & 11.6 & 6.8 & 9.2 & & 29.4 & 11.1 & 4.6 & 1.7 & 9.4 & 23.5 & 11.6 & 6.8 & 9.2\\
            \midrule
            \modelName \textsubscript{OPT} (ours)  & 44.8 & 25.2 & 13.9 & \textbf{6.8} & \textbf{12.3} & \textbf{33.1} & \textbf{28.1} & \textbf{8.6} & \textbf{18.3} &  & \textbf{33.0} & \textbf{14.8} & \textbf{6.5} & \textbf{2.9} & \textbf{9.4} & \textbf{25.4} & \textbf{14.0} & \textbf{5.3} & \textbf{9.7}\\ 
            \bottomrule
        \end{tabular}
    }
    \captionof{table}{\modelName model ablations on AudioCaps~\cite{kim2019audiocaps} and Clotho~\cite{drossos2020clotho}.}%
    \label{tab:ablations}
\end{minipage}
\vspace{-1em}
\end{figure}

\mypara{Qualitative results.}
In~\Cref{fig:qualitative examples} we can observe that our generated audio captions capture the audible events from given audio clips across a large variety of settings.
For the failure case (right), our model does not provide a very detailed caption and thus fails to capture the full content of the audio clip.

\mypara{Ablations.} We ablate the core components of our approach in \Cref{tab:ablations}. Specifically, we study the effects of our guiding components and of using different LLMs\@. Using \emph{no keywords} drastically reduces performance across all metrics. This indicates that it is important to condition the LLM on audible concepts.
When using \emph{no guiding with audio-relevancy} (i.e.\ $\beta=0$), performance degrades. Combined, this shows that both guiding techniques contribute to the overall performance of \modelName.
Since~\cite{shaharabany2023zero} uses GPT-2 as their base LLM, we also run our model with GPT-2 and still outperform their approach, even though their audio-language model~\cite{Girdhar2023ImageBindOE} is stronger than ours.

\section{Conclusion}
We introduced a zero-shot audio captioning framework that converts audio clips into textual captions using a twofold audio-based guiding approach without any training. Our proposed \modelName framework achieves state-of-the-art results on the AudioCaps and Clotho benchmarks. Furthermore, our results show that keyword-based guiding is highly beneficial for obtaining better audio captions.

\section{Acknowledgements}
The authors thank IMPRS-IS for supporting Leonard Salewski. This work was partially funded by the BMBF Tübingen AI Center (FKZ: 01IS18039A), DFG (EXC number 2064/1 – Project number 390727645), and ERC (853489-DEXIM).

\bibliographystyle{plainnat}
\bibliography{refs.bib}


\end{document}